\documentclass[12pt]{iopart}
\usepackage{color} 
\usepackage{graphicx}
\expandafter\let\csname equation*\endcsname\relax
\expandafter\let\csname endequation*\endcsname\relax
\usepackage{amsmath}
\usepackage{cite}

\begin{document}

\title[Scaling theory for two-dimensional single domain growth]
{Scaling theory for two-dimensional single domain growth driven by attachment of diffusing adsorbates}

\author{Kazuhiko Seki}

\address{Nanomaterials Research Institute(NMRI), 
National Institute of Advanced Industrial Science and Technology (AIST), 
AIST Tsukuba Central 5, Higashi 1-1-1, Tsukuba, Ibaraki 305-8565, Japan}
\ead{k-seki@aist.go.jp}
\vspace{10pt}

\begin{abstract}
Epitaxial growth methods are a key technology used in producing large-area thin films on substrates but as a result of various factors controlling growth processes the rational optimization of growth conditions is rather difficult. Mathematical modeling is one approach used in studying the effects of controlling factors on domain growth. The present study is motivated by a recently found scaling relation between the domain radius and time for chemical vapor deposition of graphene. Mathematically, we need to solve the Stefan problem; when the boundary moves, its position should be determined separately from the boundary conditions needed to obtain the spatial profile of diffusing adsorbates. We derive a closed equation for the growth rate constant defined as the domain area divided by the time duration. We obtain approximate analytical expressions for the growth rate; the growth rate constant is expressed as a function of the two-dimensional diffusion constant and the rate constant for the attachment of adsorbates to the solid domain. In experiments, the area is decreased by stopping the source gas flow. The rate of decrease of the area is obtained from theory. The theoretical results presented provide a foundation to study controlling factors for domain growth.
\end{abstract}

%

\section{Introduction}
\label{sec:I}
A key technology in producing large-area thin films on substrates is the epitaxial growth method. These epitaxial films can be single crystals with low defect densities. 
High-quality large-area films can be produced by this method. \cite{Li_09,Ago_10,Petrone_12,Chen_15,Seah_14,Tetlow_14,Feng_19}
Although epitaxial growth has been widely used, the rational optimization of growth conditions is rather difficult because of various factors controlling the growth process. \cite{Seah_14,Feng_19}
Mathematical modeling appears as one means to study the effects of controlling factors on domain growth. \cite{Tetlow_14,Kim_13,Weinan2001}

One important controlling factor on the domain growth is the diffusion of adsorbates on the two-dimensional substrate.
Adsorbates (adatoms or atom clusters) attach by diffusion to an island of solid compact domain.
The compact domain grows under two-dimensional diffusion of the adsorbates. 
However, developing an analytical theory describing the compact domain growth even  without changing morphology is challenging because we need to consider in a consistent manner a moving boundary of the solid domain and the two-dimensional time-dependent profile of the diffusing adsorbates influenced by the attachment of adsorbates to the solid domain.
In addition to the boundary conditions needed to solve the diffusion equation for adsorbates on the substrate, the position of the boundary must also be determined when it moves.
Mathematically, the extra condition is called the Stefan condition. \cite{crank_87,Krapivsky_12,Larralde_93,Burlatsky_96,Oshanin_98,PESHEVA_02,Gupta_18}
If diffusion occurs inside a circular domain and the reaction occurs at the periphery of this domain, the solution to the Stefan problem in two dimensions is given by a scaling variable determined from the conservation of the total number of adsorbates on the substrate including those incorporated in the solid domain. \cite{Krapivsky_12,Larralde_93,Forsberg_08,Sanchez_99,Havlin_95,Larralde_92,Abraham_02,Burlatsky_96}
This situation is converse to the present case for which the diffusing adsorbates on the substrate are present outside the circular solid domain.
We formulate and solve the Stefan problem for the latter situation.

Specifically, the present study is motivated by a scaling relation between the domain radius and time found from experiments on the chemical vapor deposition of graphene. \cite{Kim_12,Terasawa2015,KATO_2016}
If the radius of the domain grown during the time period $t$ is denoted by $R(t)$, the scaling relation can be expressed as $R(t)/t$ or $R^2(t)/t$ depending on the time regime. \cite{Terasawa2015}
By phenomenological fitting of the experimental data, the domain growth rate is introduced using the $R(t)/t$ relation but the data can be also described by $R^2(t)/t$ for longer time periods. 
This result reflects the difficulty in distinguishing the two behaviors in the experimental data.
From theory, the scaling relation may be expressed as $R^2(t)/t$ being constant when the circular domain growth is driven by a localized source. \cite{Larralde_93,Krapivsky_12}
The theory must be modified to allow for non-localized sources when studying the domain growth under chemical vapor deposition.

The scaling relation for $R^2(t)/t$ implies that the growth rate in the domain area is constant.
Even though the linear relation is observed between the domain area and the time period, the mechanism underlying this relationship and the factors controlling the proportionality coefficient is unclear.
The constant may depend on the two-dimensional diffusion constant and the rate of attachment of diffusing adsorbates on the substrate to the solid domain.
By properly defining the domain growth rate and relating the growth rate constant with these physical parameters, the activation energy of the physical parameter may be estimated by varying the temperature.
In our study, we also considered the rate of deposition of atoms on the two-dimensional surface and desorption of adsorbates on the two-dimensional substrate into the space above/below the surface (Figure~\ref{fig:scheme}).
Desorption is essential in removing the divergence of the concentration of adsorbates on the substrate obtained after solving the two-dimensional diffusion equation. \cite{Freeman_83}

The boundary conditions for describing the attachment of adsorbates on the substrate to the solid domain at the moving boundary is derived from the mass conservation relation and the two-dimensional diffusion equation.
The lowering of the adsorbate density at the periphery of the domain and the increase in the domain size, both arising from the attachment of adsorbates to the solid domain, are taken into account in the boundary conditions imposed on the moving boundary.
Using the boundary conditions, we obtain analytical expressions for the growth rate in terms of a two-dimensional diffusion constant and the rate of attachment of the diffusing adsorbates on the substrate.

In general, sigmoidal domain growth follows an induction period of nucleation; 
domains initially grow in isolation from other domains, and 
domain growth slows down by diffusive interaction for diffusing adsorbates among domains. \cite{Kim_12,Terasawa2015,Eres_14,Wu_16,Traytak_07}
We consider domain growth under the isolated condition from other domains. 
The scaling solution obtained in this work could be regarded as the asymptotic growth law of 
isolated domains until diffusive interaction becomes significant, and thus, slows down the domain growth. 
\section{Two-dimensional domain growth driven by diffusion}
\label{sec:II}
 \begin{figure}[h]
  \begin{center}
    \includegraphics[width=0.5\textwidth]{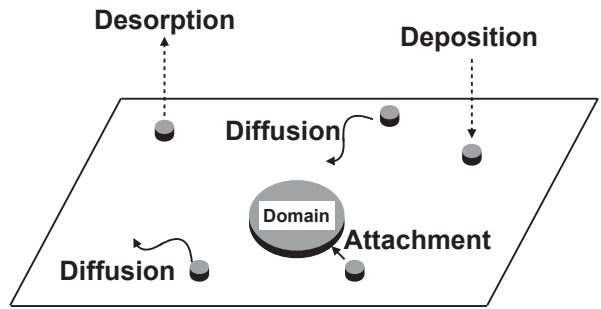}
  \end{center}
  \caption{Schematic of the circular domain growth by attachment of adsorbates (or small atom clusters). Desorption and deposition of adsorbates on the two-dimensional solid substrate are taken into account.
  }
  \label{fig:scheme}
\end{figure}
We consider the growth of a circular solid domain of radius $R(t)$ 
through the attachment of atoms or small atom clusters.  
The two-dimensional density of the solid domain is 
 denoted by $\rho$.  
 An example is the growth of a domain of graphene by adsorption of C atoms or small C atom clusters.
Hereafter, we do not distinguish between adatoms and small clusters; C atoms can be substituted by small C atom clusters if the latter is experimentally more relevant.
We consider instances when the concentration gradient toward a circular solid domain is developed by attachment of diffusing adsorbates.
We assume an isotropic concentration profile.
The diffusion flow rate of adsorbates from the circular region of radius $R_o$ is denoted by $J$.
Denoting the concentration of adsorbates on the two-dimensional substrate at distance $r$ from the center of a circular solid domain at time $t$ by $C(r,t)$, $J$ is then expressed as
\begin{align}
J=2 \pi R_o D \left. \frac{\partial}{\partial r}C(r,t) \right|_{r=R_o},
\label{eq:inflow}
\end{align}
where $D$ denotes the two-dimensional diffusion constant of adsorbates on the substrate.
In the presence of desorption, we prove that $J$ approaches zero as $R_o \rightarrow \infty$.
We consider the circumstance when the concentration of the adsorbates on the substrate increases by deposition of atoms from a three-dimensional phase above and/or below the two-dimensional surface; adsorbates impinge from either the gas phase (by deposition and decomposition of clusters) or through the substrate to the surface. \cite{Taira_2017,McCarty_09,Losurdo_11,Celebi_13}
The periphery of the circular solid domain grows by attachment of adsorbates at $R(t)$ 
(see figure~\ref{fig:scheme}).
The continuity equation for the concentration of adsorbates surrounding the circular solid domain is expressed as
\begin{align}
\frac{\partial}{\partial t} C(r,t)
=\frac{1}{r} \frac{\partial}{\partial r} D r\frac{\partial}{\partial r} C(r,t)-k_d C(r,t)+g.
\label{eq:continuity}
\end{align}
where the deposition rate of adsorbates from the  three-dimensional phase to the two-dimensional surface per unit area is denoted by $g$, and the desorption rate of adsorbates per unit area of the substrate to the three-dimensional phase is denoted by $k_d$.

The increase in $R(t)$ through the attachment of adsorbates must be taken into account consistently by relating the rate of increase in $R(t)$ and the rate of attachment of adsorbates at the periphery of the solid domain obeying the continuity equation given by equation~(\ref{eq:continuity}).
As the boundary moves, its position must be determined by a self-consistent condition, which is constructed from the mass conservation law inside the region between radii $R_o$ and $R(t)$; specifically,
\begin{align}
\int_{R(t)}^{R_o} 2 \pi r dr\, C(r,t)+\rho \pi R^2(t)=\int_{R(t)}^{R_o} 2 \pi r dr\, gt-\int_{R(t)}^{R_o}dr\, 2\pi r \int_0^t dt_1 k_d C(r,t_1)+\int_0^t dt_1J(t_1),
\label{eq:cons}
\end{align}
where we omit a constant on the right-hand side of equation~(\ref{eq:cons}) that arises from the arbitrariness of the initial time.
By differentiating both sides by $t$, we obtain
\begin{multline}
\int_{R(t)}^{R_o}dr\, 2\pi r \frac{\partial}{\partial t} C
+2\pi R(t)\left[\rho-C(R(t),t)+gt\right]\frac{\partial R(t)}{\partial t}\\
=\pi g \left[ R_o^2-R^2(t)\right]-\int_{R(t)}^{R_o}dr\, 2\pi r k_d C(r,t)+J.
\label{eq:consdiff_1}
\end{multline}
We substitute equation~(\ref{eq:continuity}) into equation~(\ref{eq:consdiff_1}), perform the spatial integration to obtain
\begin{align}
2\pi R(t)\left(\rho- C(R(t),t)+gt\right)\frac{\partial R}{\partial t}
=\left. 2\pi D R(t) \frac{\partial C}{\partial r} \right|_{r=R(t)}.
\label{eq: consdiffo}
\end{align}
In equation~(\ref{eq: consdiffo}), $gt$ signifies the homogeneous accumulation of adsorbates on the substrate by adsorbate deposition from the three-dimensional phase.
We assume that the concentration of accumulated adsorbates on the substrate is small compared with $\rho$ and $C(R(t),t)$ and is ignored in the following analysis; {\it i.e.}, we consider
\begin{align}
\left[\rho- C(R(t),t) \right] \frac{\partial R}{\partial t}
=\left. D \frac{\partial C}{\partial r} \right|_{r=R(t)},
\label{eq:bcC1}
\end{align}
which is the Stefan boundary condition, where the boundary position is determined to satisfy the mass conservation law.
In the above, we assumed implicitly that the rate of attachment of adsorbates to the circular solid domain is infinitely fast.
The above formulation can be also applied for fast equilibration between the attachment of adsorbates to and release of adsorbates from the circular domain; the detailed balance between the attachment and release of adsorbates at the solid circular domain is attained so that the local equilibrium of attachment--release is almost attained at the moving boundary of the growing solid circular domain. \cite{McLean_99}
When both sides of equation~(\ref{eq:bcC1}) are multiplied by $2\pi R(t)$, 
equation~(\ref{eq:bcC1}) can be interpreted as saying that  
the number of atoms $\pi R(t)^2 \rho$ entering through the domain surface per unit time 
is given by the concentration gradient perpendicular to the domain boundary 
times the boundary length $2 \pi R(t)$. \cite{ZINKEALLMANG_92}
$C(R(t),t)$ on the left-hand side of equation (\ref{eq:bcC1}) is often 
ignored but will be kept below. \cite{ZINKEALLMANG_92}

We solve the diffusion equation given by equation~(\ref{eq:continuity}) imposing the Stefan boundary condition as well as
\begin{align}
C(R(t),t)= C_0,
\label{eq:db}
\end{align}
where $C_0$ denotes the concentration accounting for the desorption of adsorbates from the solid circular domain to the two-dimensional surface. \cite{McLean_99}

It is mathematically difficult to solve the diffusion equation given by equation~(\ref{eq:continuity}) 
under the boundary conditions given by equations~(\ref{eq:bcC1}) and (\ref{eq:db}) for all time regimes. 
We focus on the asymptotic time regime of single domain growth. 
The effect of other domains is ignored. 
We seek a scaling solution in which the time dependence of $C(r,t)$ is governed by the moving circular solid boundary denoted by $R(t)$ by assuming $C(r,t)= C(\xi)$, 
where $\xi$ is given by $\xi=r/R(t)$. \cite{Krapivsky_12,Larralde_93,Burlatsky_96,Oshanin_98}
The method has been used to solve a mathematically similar problem; growth of a wetting monolayer. \cite{Burlatsky_96,Oshanin_98}
As $(\partial \xi)/(\partial r)=1/R(t)$, we obtain
\begin{align}
\frac{\partial C}{\partial t}=-\frac{\partial R}{\partial t}
\frac{\xi}{R} \frac{\partial C}{\partial \xi},\,
\frac{\partial C}{\partial r}=\frac{1}{R} \frac{\partial C}{\partial \xi},\,
\frac{\partial^2 C}{\partial r^2}=\frac{1}{R^2} \frac{\partial^2 C}{\partial \xi^2}.
\label{eqs}
\end{align}
Equation (\ref{eq:continuity}) can be rewritten as
\begin{align}
\frac{\partial^2 C}{\partial \xi^2}+\frac{1}{\xi} \frac{\partial C}{\partial \xi}-q C
=-\frac{R}{D} \frac{\partial R}{\partial t} \xi \frac{\partial C}{\partial \xi}-g_n,
\label{eq:cont1}
\end{align}
where $g_n$ 
(dimensionless generation rate of adsorbates {\it e.g.} by deposition)
and $q$ (dimensionless desorption rate of adsorbates) are defined as
\begin{align}
g_n\equiv g R^2/D \mbox{ and } q\equiv k_d R^2/D.
\label{eq:gn}
\end{align}
Equation (\ref{eq:cont1}) is consistent if $(R/D) (\partial R)/(\partial t)$, which is in principle a function of time, is time-independent, that is, imposing \cite{Krapivsky_12,Larralde_93}
\begin{align}
\frac{R}{D} \frac{\partial R}{\partial t}= 2 \alpha.
\label{eq:consistency}
\end{align}
with $\alpha$ a constant. From this consistency condition, we have
\begin{align}
R^2(t)= 4 \alpha D t
\label{eq:c1}
\end{align}
and $\alpha$ is found using the Stefan boundary condition, equation~(\ref{eq:bcC1}), when the diffusion equation is solved using the boundary conditions.

By requiring scaling, equation~(\ref{eq:cont1}) may be rewritten as
\begin{align}
\frac{\partial^2 C}{\partial \xi^2}+\left( \frac{1}{\xi}+2 \alpha \xi \right)
\frac{\partial C}{\partial \xi}-q C=-g_n.
\label{eq:cxides}
\end{align}
On introducing a new variable, $z=-\alpha \xi^2$, equation~(\ref{eq:cxides}) transforms to
\begin{align}
z\frac{\partial^2 C}{\partial z^2}+\left( 1-z\right)
\frac{\partial C}{\partial z}+\frac{q}{4\alpha} C= \frac{g_n}{4\alpha}.
\label{eq:cxidez}
\end{align}
The homogeneous part of equation~(\ref{eq:cxides}) is Kummer's equation. \cite{NIST}
We choose the non-diverging solution when $\xi \rightarrow \infty$; specifically,
\begin{align}
C(\xi)=C_1 f(\xi)+(g_n/q),
\label{solC:desp}
\end{align}
where $C_1$ is the constant of integration to be determined from the boundary condition, and $f(\xi)$ is given by
\begin{align}
f(\xi)=\exp\left(-\alpha \xi^2\right) U\left(1+\frac{q}{4\alpha},1,\alpha \xi^2\right),
\label{eq:fU}
\end{align}
with $U(a,b,z)$ denoting Kummer's confluent hypergeometric function of the second kind, \cite{NIST} and $\alpha$ determined from the Stefan boundary condition.

Using the boundary condition for the concentration profile given by equation~(\ref{eq:db}), we find
\begin{align}
C_1=\left[C_0-(g_n/q)\right]/f(1).
\label{eq:C1desp_inf}
\end{align}
The Stefan boundary condition given by equation~(\ref{eq:bcC1}) is expressed as
\begin{align}
\left. 2 \alpha[\rho-C(1)]-\frac{\partial C}{\partial \xi}\right|_{\xi=1}=0,
\label{eq:Stefan1}
\end{align}
and we obtain
\begin{align}
\rho-(g_n/q)=\left[f'(1)/(2\alpha)+f(1) \right]C_1.
\label{eq:Stefan1n}
\end{align}
By eliminating $C_1$ from equations~(\ref{eq:C1desp_inf}) and (\ref{eq:Stefan1n}), the closed equation for $\alpha$ becomes
\begin{align}
f(1)\left(\rho-C_0\right)=-\frac{f'(1)}{2\alpha}\left[(g_n/q)-C_0\right].
\label{eq:alpha_des8}
\end{align}

An explicit expression for $f'(1)$ is obtained with
\begin{align}
f'(1)=-2 \exp\left(-\alpha \right) U\left(q/(4\alpha),0,\alpha \right),
\label{eq:f11}
\end{align}
where we have used $f'(\xi)=-2\exp\left(-\alpha \xi^2\right) U\left(q/(4\alpha),0,\alpha \xi^2\right)/\xi$.\cite{NIST}
We find from equation~(\ref{eq:fU})
\begin{align}
f(1)=\exp\left(-\alpha \right) U\left(1+\frac{q}{4\alpha},1,\alpha \right).
\label{eq:fU1}
\end{align}
Using equations~(\ref{eq:f11}) and (\ref{eq:fU1}), equation~(\ref{eq:alpha_des8}) is finally expressed as
\begin{align}
\alpha \left(\rho-C_0\right) U\left(1+\frac{q}{4\alpha},1,\alpha \right)=\left[(g_n/q)-C_0\right] U\left(q/(4\alpha),0,\alpha \right), 
\label{eq:alpha_des9}
\end{align}
which is the first main result of this paper.

We study an instance when the growth rate of the solid domain is much smaller than the rate of adsorbate diffusion.
We simplify equation~(\ref{eq:alpha_des9}) given $\alpha <1$.
We also assume a long adsorbate lifetime in the absence of a solid domain; the condition is expressed as $q=k_d R^2/D<1$.
We introduce an approximation for Kummer's confluent hypergeometric function of the second kind given in \cite{NIST}
\begin{align}
U(a,1,z) &\approx- \frac{\psi (a)+\ln(z)+2 \gamma}{\Gamma (a)},
\label{eq:appU1}\\
U(a,0,z) &\approx \frac{1}{\Gamma(a+1)},
\label{eq:appU0}
\end{align}
where $\Gamma (a)$, $\psi (a)$, and $\gamma$ denote the gamma function, psi (digamma) function, and Euler's constant, respectively.
By substituting equations~(\ref{eq:appU1}) and (\ref{eq:appU0}), equation~(\ref{eq:alpha_des9}) simplifies,
\begin{align}
\alpha \approx- \frac{(g_n/q )-C_0}{\rho-C_0}
\frac{1}{\psi [1+q/(4\alpha) ]+\ln(\alpha)+2 \gamma}.
\label{eq:alphasimple}
\end{align}
Using the asymptotic expansion of psi function, $\psi[1+q/(4\alpha)]\approx\ln[q/(4\alpha) ]$, \cite{NIST} we have
$\psi [1+q/(4\alpha) ]+\ln(\alpha) \approx \ln[q/(4\alpha)]+\ln (\alpha) = \ln(q/4) $.
Finally, we obtain
\begin{align}
\alpha \approx \frac{(g/k_d)-C_0}{\rho-C_0} \frac{1}{\ln[4D/(k_d R^2)]-2\gamma},
\label{eq:main}
\end{align}
where $\gamma=0.577 \cdots$.
With this approximation, $R^2(t)$ given by equation~(\ref{eq:c1}) is expressed as
\begin{align}
R^2(t)\approx \frac{4 D t}{\ln[4D/(k_d R^2)]-2\gamma}
\left(\frac{(g/k_d)-C_0}{\rho-C_0}\right),
\label{eq:mainR2}
\end{align}
which is the second main result of this paper.
When the rate of adsorbate attachment to the solid phase at the periphery of the solid domain is sufficiently fast compared with the rate of its reverse process, we may set $C_0=0$.
 \begin{figure}[h]
  \begin{center}
    \includegraphics[width=0.5\textwidth]{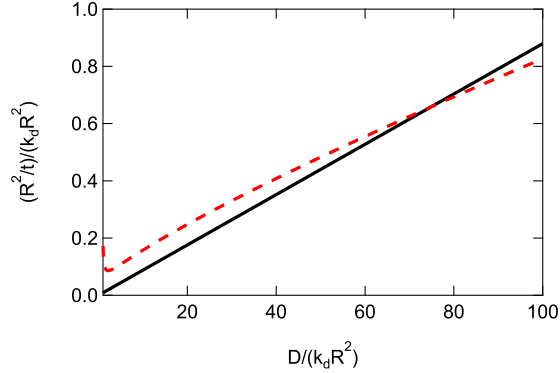}
  \end{center}
  \caption{(Color online) Dimensionless $R(t)^2/t\mbox{ } (4\alpha/q)$ is plotted against $D\mbox{ } (1/q)$ when $[(g/k_d)-C_0]/(\rho-C_0)=0.01$. The dimensionless quantities are obtained using $k_d R^2$. The thick (black) solid line indicates the numerical exact results obtained by solving equation~(\ref{eq:alpha_des9}); the red dashed line indicates the approximate results from equation~(\ref{eq:mainR2}).
  }
  \label{fig:SD_r_0p01}
\end{figure}

\begin{figure}[h]
  \begin{center}
    \includegraphics[width=0.5\textwidth]{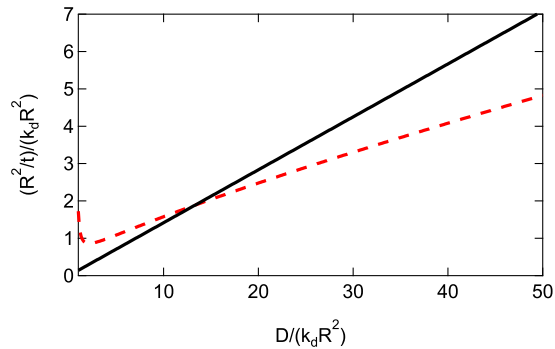}
  \end{center}
  \caption{(Color online) Dimensionless $R(t)^2/t\mbox{ } (4\alpha/q)$ is plotted against $D\mbox{ } (1/q)$ when $[(g/k_d)-C_0]/(\rho-C_0)=0.1$. The dimensionless quantities are obtained using $k_d R^2$. The thick (black) solid line indicates the numerical exact results obtained by solving equation~(\ref{eq:alpha_des9}); the red dashed line indicates the approximate results from equation~(\ref{eq:mainR2}).
  }
  \label{fig:SD_r_0p1}
\end{figure}

In figures~\ref{fig:SD_r_0p01} and \ref{fig:SD_r_0p1}, $R(t)^2/t$ calculated from $4\alpha D$ is shown as a function of the values of the diffusion constant.
Both quantities are expressed in dimensionless form using $k_d R^2$.
The approximate expression given by equation~(\ref{eq:mainR2}) is close to the exact numerical results when $[(g/k_d)-C_0]/(\rho-C_0)=0.01$ compared with that obtained for $[(g/k_d)-C_0]/(\rho-C_0)=0.1$.
The approximated expression is obtained by assuming that $\alpha$ is small.
In zeroth-order approximation, $\alpha$ is proportional to $[(g/k_d)-C_0]/(\rho-C_0)$.
Therefore, the deviation of the approximate results from the exact results increases as the value of $[(g/k_d)-C_0]/(\rho-C_0)$ increases.
In addition to the condition for $\alpha$, the approximate expression given by equation~(\ref{eq:mainR2}) is derived by assuming that $k_d R^2/D$ is small.
As a consequence, the approximate results deviate from the exact numerical results when $D/k_d R^2 <1$ regardless of the value of $\alpha$.

\section{Finite attachment rate}
\label{sec:III}
So far, we have assumed that adsorbate attachment occurs infinitely fast.
The above results are valid when the attachment rate is faster than any other time scale in the system.
In this section, we generalize the above results for a finite rate of attachment at the periphery of the solid domain.
The rate of attachment of adsorbates on the substrate to the solid phase is denoted by $k$.
Here, the adsorbates located on the periphery of the solid domain do not necessary attach to the domain and may bounce back in contrast to our previous assumption that the adsorbates at the boundary either attach rapidly to the solid domain or equilibrated rapidly at the boundary.
In theory, we may assume reflecting boundary conditions at the domain boundary and consider the reaction term with the rate of attachment per unit area denoted by $k$ as
\begin{align}
\frac{\partial}{\partial t} C(r,t)
=\frac{1}{r} \frac{\partial}{\partial r} D r\frac{\partial}{\partial r} C(r,t)
- k \frac{C(r,t)}{2\pi r} \delta(r-R(t)-\epsilon/2)+g-k_d C(r,t),
\label{eq:reaction_d}
\end{align}
where $\epsilon$ is a small constant; the limit $\epsilon \rightarrow 0$ is taken later.
The growth rate of the domain area should be proportional to the rate of attachment given by $k C(R(t)+\epsilon/2,t)$.

An alternative theoretical method is possible using the diffusion equation given by equation~(\ref{eq:continuity}) without the term describing attachment; the attachment is taken into account by the boundary condition.
The boundary condition for equation~(\ref{eq:continuity}) is obtained using equation~(\ref{eq:reaction_d}) with the reflecting boundary condition at the solid domain boundary.
By multiplying $2 \pi r$ and integrating equation~(\ref{eq:reaction_d}), we obtain
\begin{multline}
\int_{R(t)}^{R(t)+\epsilon} 2 \pi r dr
\frac{\partial}{\partial t} C(r,t)
=\left. 2 \pi D r\frac{\partial}{\partial r} C(r,t) \right|_{r=R(t)+\epsilon}
- k C(R(t)+\epsilon/2,t)+\\
\int_{R(t)}^{R(t)+\epsilon} 2 \pi r dr\left[g-k_d C(r,t)\right],
\end{multline}
where the reflecting boundary condition given by $\partial C/(\partial r) |_{r=R(t)}=0$ is used.
In the limit $\epsilon \rightarrow 0$, the above equation yields
\begin{align}
\left. \frac{\partial}{\partial r} C(r,t) \right|_{r=R(t)}
&=\frac{k}{2 \pi R(t)D} C(R(t),t).
\label{eq:intcontd}
\end{align}
In the presence of a back-reaction for detachment, equation~(\ref{eq:intcontd}) generalizes, \cite{Schwoebel_69,McLean_99}
\begin{align}
\left. \frac{\partial}{\partial r} C(r,t) \right|_{r=R(t)}
&=\frac{k}{2 \pi R(t)D} \left[C(R(t),t)-C_0\right],
\label{eq:intcontd1}
\end{align}
where we postulate that the current of adsorbates at the solid domain boundary is linear in the deviation of the density of adsorbates at the periphery of the solid domain from their concentration when in equilibrium for the flat surface. \cite{McLean_99}
We ignored the Gibbs--Thomson effect, \cite{ZINKEALLMANG_92}
which is taken into account by changing $C_0$ to $C_0+a/R(t)$, where $a$ is a positive constant. \cite{Ohta}

In addition to the boundary condition for the concentration profile of adsorbates, the Stefan boundary condition is required to describe the growth of the solid area.
The Stefan boundary condition can be derived from equation~(\ref{eq:consdiff_1}) for a finite attachment rate 
as shown below.
We substitute equation~(\ref{eq:reaction_d}) into equation~(\ref{eq:consdiff_1}) and perform a spatial integration using the perfectly reflecting boundary condition at $r=R(t)$.
By taking the limit $\epsilon \rightarrow 0$, we obtain 
\begin{align}
2\pi R(t)\left(\rho- C(R(t),t)-gt\right)\frac{\partial R}{\partial t}=
k\left( C(R(t),t)-C_0 \right),
\label{eq:consdiff0d}
\end{align}
where $C_0$ is zero in the absence of a back reaction for detachment.
By combining equations~(\ref{eq:intcontd1}) and (\ref{eq:consdiff0d}), we note that the Stefan boundary condition is given by equation~(\ref{eq:bcC1}) and therefore holds even when the rate of attachment is finite.

We solve the diffusion equation given by equation~(\ref{eq:cont1}) expressed using the scaling variable.
The boundary condition is next written in terms of the scaling variable,
\begin{align}
\left. \frac{\partial C}{\partial \xi}\right|_{\xi=1}=\frac{k}{2\pi D}
\left(
C(1)-C_0 \right).
\label{eq:Stefanp1}
\end{align}
The Stefan boundary condition is unchanged and is given by equation~(\ref{eq:Stefan1}).
Substituting equation~(\ref{solC:desp}) [$C(\xi)=C_1 f(\xi)+(g_n/q)$] into equation~(\ref{eq:Stefanp1}) yields
\begin{align}
C_1=\frac{C_0-(g_n/q)}
{f(1)-(2\pi D/k)f'(1)}.
\label{eq:C1desp}
\end{align}
By eliminating $C_1$ from equations~(\ref{eq:C1desp}) and (\ref{eq:Stefan1n}), a closed equation for $\alpha$ is obtained,
\begin{align}
\alpha \left[\rho-C_0\right] U\left(1+\frac{q}{4\alpha},1,\alpha \right)
 =\left[\frac{g_n}{q}-C_0-\frac{4\pi\alpha D}{k}
 \left(\rho-\frac{g_n}{q} \right)\right] U\left(\frac{q}{4\alpha},0,\alpha \right).
\label{eq:alpha_des9p}
\end{align}
Substituting the approximation for Kummer's confluent hypergeometric function given by equations~(\ref{eq:appU1}) and (\ref{eq:appU0}), together with the asymptotic expansion of the psi function given by $\psi[1+q/(4\alpha)] \approx \ln[q/(4\alpha) ]$, \cite{NIST} simplifies the expression for $R^2(t)$,
\begin{align}
R^2(t)&\approx 
\frac{4 D t}{\ln[4D/(k_d R^2)]-2\gamma+4\pi D(\rho-g/k_d)/[k(\rho-C_0)]}
\left(\frac{(g/k_d)-C_0}{\rho-C_0}\right)
\label{eq:mainR2pdo} \\
&\approx \frac{4 D t}{\ln[4D/(k_d R^2)]-2\gamma+4\pi D/k}
\left(\frac{(g/k_d)-C_0}{\rho-C_0}\right).
\label{eq:mainR2pd}
\end{align}
Equations (\ref{eq:alpha_des9p}) and (\ref{eq:mainR2pd}) are the main results of this section.
 \begin{figure}[h]
 \begin{center}
    \includegraphics[width=0.5\textwidth]{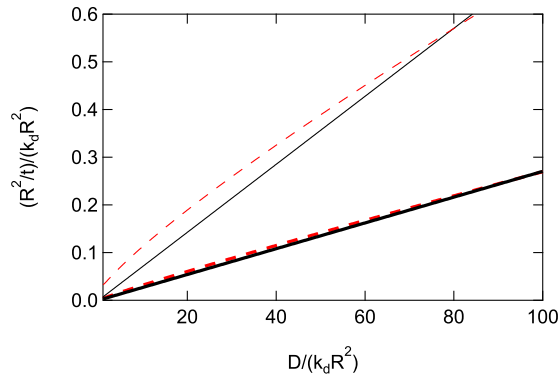}
  \end{center}
  \caption{(Color online) Dimensionless $R(t)^2/t\mbox{ } (4\alpha/q)$ is plotted against $D\mbox{ } (1/q)$ when $[(g/k_d)-C_0]/(\rho-C_0)=0.01$. The dimensionless quantities are obtained using $k_d R^2$. The thin and thick lines distinguish results with $k(\rho-C_0)/[4\pi D(\rho-g_n/q)]=1$ and $0.1$, respectively. The (black) solid line indicates the numerical exact results obtained by solving equation~(\ref{eq:alpha_des9p}); the (red) dashed line indicates the approximate results from equation~(\ref{eq:mainR2pdo}).
  }
  \label{fig:SD_r_0p01d}
\end{figure}

In figure~{\ref{fig:SD_r_0p01d}, we compare the approximate results obtained from equation~(\ref{eq:mainR2pd}) with the exact numerical results obtained from equation~(\ref{eq:alpha_des9p}) using $R^2(t)=4\alpha Dt$.
Apart from $k(\rho-C_0)/(4\pi D)$, the parameters are the same as those appearing in figure~\ref{fig:SD_r_0p01}.
We note that the approximate results obtained using equation~(\ref{eq:mainR2pd}) are close to the exact numerical results.
A better agreement was found by decreasing the value of the attachment rate to the solid domain.
As long as the results obtained in the limit $k \rightarrow \infty$ reproduce the exact numerical results, the approximate results obtained using equation~(\ref{eq:mainR2pd}) reproduce the exact numerical results.
In this sense, the effect of a finite rate of attachment is well taken into account by equation~(\ref{eq:mainR2pd}).
\section{Comparison with experiments}
\label{sec:IV}
Given the experimental uncertainty, whether the area or the square root of the area is proportional to time is still not clear. \cite{Terasawa2015}
The square root of the area is proportional to time 
if the domain area growth rate ($\partial \pi R(t)^2/\partial t$) is proportional to the peripheral length.
Growth then proceeds under a constant concentration of adsorbates. \cite{Terasawa2015}
The Stefan boundary condition given by equation~(\ref{eq: consdiffo}) indicates 
that the domain area growth rate is 
proportional to the peripheral length times the concentration gradient at the domain boundary. 
When the square root of the area is proportional to time,  the domain area growth rate 
is proportional to the peripheral length; 
growth then proceeds under a constant concentration gradient of adsorbates at the domain boundary.\cite{Terasawa2015,Huijun_15}
The latter assumption is incorrect for circular domain growth, and the concentration gradient at the domain boundary is inversely proportional to the domain radius.
Because the peripheral length is proportional to the domain radius, 
the domain area growth rate becomes independent of the domain radius.
As a result, the domain area is proportional to time according to the Stefan boundary condition.

In experiments on the growth of graphene domains (on which this paper focuses), 
the growth rate is as small as $1 \mu \mbox{m}^2/$~s. \cite{Terasawa2015,Kim_12}
The atomic area density of graphene is given by $\rho=3.82 \times 10^{19}$~m$^{-2}$ and the surface carbon adatom concentration is roughly $1 \times 10^{16}$~m$^{-2}$.
In equation~(\ref{eq:mainR2}), $g/k_d$ is identified as the equilibrium surface adatom concentration and $[(g/k_d)-C_0]/\left(\rho-C_0\right)$ estimated to be $10^{-3}$.
When the two-dimensional diffusion constant is $10^{-6}$~m$^2$/s, the right-hand side of equation~(\ref{eq:mainR2}) yields an estimate $10^{-9}$~m$^2$/s whereas the left-hand side of equation~(\ref{eq:mainR2}) is $10^{-12}$~m$^2$/s.
The result indicates that the growth rate may be limited by the rate of attachment rather than diffusion.
By taking the limit of $4\pi D/k \gg \ln[4D/(k_d R^2)]-2\gamma$, the reaction-limited expression obtained from equation~(\ref{eq:mainR2pd}) is
\begin{align}
R^2(t)\approx \frac{k }{\pi}\frac{(g/k_d)-C_0}{\rho} t.
\label{eq:mainR2pdr}
\end{align}
In experiments, the temperature dependence of the growth rate is examined assuming the rate given by $R(t)^2/t$. \cite{Kim_12}
By applying the Arrhenius plot to the domain growth rate, the activation energy is estimated to be $2.6$~eV. \cite{Kim_12}
In accordance with equation~(\ref{eq:mainR2pdr}), the activation energy corresponding to $g/k_d$ as well as $k$ should be taken into account.
We may decouple these contributions by changing the flow rate of the source gas at each temperature. \cite{Terasawa2015}
Careful examination of the growth rate given by $R^2(t)/t$ is required to study experimentally the reaction-limited growth.

In equation~(\ref{eq:mainR2pdr}), $R^2(t)$ decreases when $(g/k_d)-C_0$ is negative.
$(g/k_d)-C_0$ could be negative if the deposition of atoms to the two-dimensional surface is absent while the desorption from the two-dimensional surface is present, $g/k_d=0$.
Indeed, $R^2(t)$ decreases by stopping the source gas flow. \cite{Terasawa2015}
In experiments, the rate of decrease of $R^2(t)$ was smaller than the rate of increase of $R^2(t)$. \cite{Terasawa2015}
$C_0$ represents the concentration just outside the solid domain when detailed balance is attained at the periphery of the solid domain; we may assume that $C_0$ is the same 
regardless of the direction in which the domain boundary moves.
The experimental situation of the smaller decreasing rate compared with the increasing rate of the solid domain can be obtained when $C_0<g/(2 k_d)$, where $g$ and $k_d$ denote the constants for the deposition and desorption rates under the source gas flow, respectively.
The theoretical prediction can be studied experimentally if $C_0$ can be measured.
Although the experimental determination of $C_0$ has not yet been performed and could be difficult, our results are consistent with the fact that the direction in which the boundary moves can be reversed by stopping the source gas flow so that $g/k_d=0$. \cite{Terasawa2015}
\section{Summary}
\label{sec:V}
We have studied the growth of a solid domain on a two-dimensional substrate from attachment of adatoms or small clusters.
The growth is driven by deposition of adsorbates from the three-dimensional phase onto the two-dimensional substrate and the growth rate was determined subject to mass conservation on the substrate; the boundary position was determined so that mass conservation is fulfilled.
First, we assumed that detailed balance between attachment and release of adsorbates at the periphery of the solid domain is attained fast enough compared with other time scales describing diffusion and boundary motion.
Then, the results were generalized to take into account a finite rate of attachment of adsorbates onto the solid domain.

The Stefan problem in two dimensions was formulated by taking into account desorption of adsorbates into the three-dimensional phase from the two-dimensional substrate.
If desorption is ignored, the concentration profile of adsorbates becomes infinite in the limit $r \rightarrow \infty$.
To avoid the divergence, the desorption of adsorbates must be taken into account.
Otherwise, a screening length should be introduced phenomenologically to avoid the divergence. \cite{ZINKEALLMANG_92}
We explicitly took into account desorption and deposition of adsorbates 
and considered in a consistent manner a moving boundary of the solid domain and the concentration profile.  

The growth rate constant, {\it i.e.}, domain area divided by elapsed time, obeys a closed equation given by Kummer's confluent hypergeometric function of the second kind [equations~(\ref{eq:alpha_des9}) and (\ref{eq:alpha_des9p})].
We obtained, after applying an approximation, analytical expressions for the growth rate given by equations~(\ref{eq:mainR2}), (\ref{eq:mainR2pd}), and equation~(\ref{eq:mainR2pdr}). 
Here, we rewrite these expressions in terms of the degree of supersaturation of the two-dimensional surface as it is more physically relevant.
$g/k_d$ can be interpreted as the concentration of adsorbates located far from the domain, where the uniform concentration is maintained; we express it as $C_\infty=g/k_d$.
In contrast, $C_0$ indicates the concentration at the periphery of the domain when the domain size is unchanged because the rate of attachment of the adsorbates to the domain equals the rate of release of adsorbates from the domain; thus, $C_0$ can be regarded as the equilibrium surface concentration of adsorbates at the periphery of the domain.
The degree of supersaturation of the two-dimensional surface is defined by $\sigma=(C_\infty-C_0)/C_0$.
Using these quantities, equation~(\ref{eq:mainR2pd}) can be rewritten as
\begin{align}
R^2(t)&\approx \frac{4 D t}{\ln[4D/(k_d R^2)]-2\gamma+4\pi D/k}
\left(\frac{C_\infty-C_0}{\rho}\right)
\label{eq:expression1}\\
&=\frac{4 D t}{\ln[4D/(k_d R^2)]-2\gamma+4\pi D/k}
\left(\frac{C_0 \sigma}{\rho}\right).
\label{eq:expression2}
\end{align}
In the diffusion-controlled limit, equation~(\ref{eq:expression2}) reduces to
\begin{align}
R^2(t)&\approx\frac{4 D t}{\ln[4D/(k_d R^2)]-2\gamma}\frac{C_0 \sigma}{\rho}.
\label{eq:expression3}
\end{align}
This is the expression corresponding to equation~(\ref{eq:mainR2}).
In the reaction-controlled limit, equation~(\ref{eq:expression2}) reduces to
\begin{align}
R^2(t) &\approx\frac{k }{\pi}\frac{C_0 \sigma}{\rho} t.
\label{eq:expression4}
\end{align}
This is the expression corresponding to equation~(\ref{eq:mainR2pdr}).

In experiments on graphene growth, the domain area was found to increase linearly with time in the asymptotic growth regime before a competing adsorption interaction occurs among domains. \cite{Kim_12,Terasawa2015,KATO_2016}
Our results are consistent with the experimental observations.
The constant associated with the growth rate was expressed as a function of the two-dimensional diffusion constant of adsorbates and the rate constant for the attachment of adsorbates to the solid domain.
In experiments, the area was decreased by stopping the source gas flow. \cite{Terasawa2015}
The rate of decrease of the area is obtained from theory. 
The theoretical results presented in this paper provide a foundation to study controlling factors for domain growth by changing deposition rates, the rate of attachment to the solid domain and the two-dimensional diffusion constant; these controlling factors can be changed in experiments by varying the temperature and source gas flow rates.
The domain growth rate depends on temperature through $D$, $\sigma$ and $g/k_d$ in the diffusion-controlled limit, 
while the domain growth rate depends on temperature through $k$, $\sigma$ and $g/k_d$ in the reaction-controlled limit.  
If the temperature dependence of $\sigma$ could be estimated from the induction period of nucleation and  $g/k_d$ from 
the temperature dependence of $C_\infty$, 
the temperature dependence of either $D$ or $k$ could be obtained from the domain growth rate. 

The growth laws obtained here are not specific to graphene, but rather are relevant to a wide range of 
materials on substrates as long as the domain can be regarded as having near-circular shape. 
Recently, transition metal dichalcogenides have attracted great attention for optoelectronic applications, which form 
the 2D monolayers by chemical vapor deposition. \cite{Li_17}
In the early stage, irregular polygons are appeared and the morphology evolves into a triangle shape. 
In a different context, relaxation of surface morphologies to a stable shape have been studied theoretically. \cite{Khare_98,Khare_96,Einstein_14} 
Strictly speaking, graphene domains also show polygon shapes such as hexagonal shape. 
The effect of morphology on the domain growth should be carefully studied both theoretically and experimentally. 
We are currently studying the growth of polygonal domains. 
 
We considered domain growth of an isolated domain until diffusive interaction slowdowns the domain growth. 
A large domain could be grown from the isolated domain and the growth law could differ from that of surface coverage. 
It should be remembered that the growth of surface coverage may not follow the behavior of single domain area growth 
owing to the time-dependence in the number-size distribution of domains. 
The surface coverage can be obtained from the domain size distribution. 
The time dependence of the domain size distribution has been studied both theoretically and experimentally but 
is beyond the scope of the present study. \cite{Stroscio_94,Bartelt_92,ZINKEALLMANG_92,Amar_94,Einstein_14,EVANS_06,Das_18}

\ack
I thank Professor Koichiro Saiki for informative discussion. 
\nocite{*}
\section*{References}
\bibliographystyle{iopart-num}

\end{document}